%% file: article.tex
\documentclass[aps,prl,twocolumn,letterpaper]{revtex4-1} 

\usepackage[T1]{fontenc}
\usepackage[english,UKenglish]{babel}
\usepackage[utf8]{inputenc}
\usepackage{latexsym} 			
\usepackage{graphicx}			
\usepackage{color}    			
\usepackage[centertags]{amsmath} 	
\usepackage{amsxtra} 			
\usepackage{amsfonts} 			
\usepackage{amssymb}  			
\usepackage{bbm}      			
\usepackage{ae,aeguill}			
\usepackage{type1cm}			
\usepackage{mathrsfs}			

\usepackage[bookmarks,			
            bookmarksopen=false,	
            bookmarksnumbered=true,	
            pdfstartview=Fit, 		
	    linkbordercolor={0 0 1},		
            citebordercolor={0 0.80 0.2},	
	    urlbordercolor={0.55 0.11 0.38},	
            colorlinks,			
            linkcolor=blue,	
	    citecolor=blue,	
            urlcolor=blue,	
	    filecolor=blue 		
            pdfauthor={Alexander Hentschel},
	    plainpages=false,
	    pdfpagelabels]{hyperref}	

\graphicspath{{./},{./Graphics/},{./../Graphics/}}

\newcommand{\hs}[1] {\hspace{#1}}
\newcommand{\vs}[1] {\vspace{#1}}

\newcommand{\tn}[1]{\textnormal{#1}}

\newcommand{\braket}[2]{\langle #1 | #2 \rangle}

\newcommand{\ket}[1]{| #1 \rangle}

\newcommand{\minfrac}[2]{\genfrac{}{}{}{1}{#1}{#2}}

\newcommand{\tphase}{\varphi}
\newcommand{\pEst}{\widetilde{\varphi}}

\newcommand{\kleinc}{\hat{c}}

\newcommand{\swarmsize}{\Xi}

\newcommand{\vac}{\ket{\tn{vac}}}

\definecolor{grey}{rgb}{0.6,0.6,0.6}
\definecolor{maroon}{rgb}{0.40,0.08,0.30}
\definecolor{darkblue}{rgb}{0.0,0.0,0.55}
\definecolor{forestgreen}{rgb}{0.13, 0.54, 0.13}
\newcommand{\SupplementMaterial}[1]{}

\renewcommand{\atop}[2]{\genfrac{}{}{0pt}{}{#1}{#2}}

\begin{document}

\renewcommand{\figurename}{\fontseries{bx}\selectfont Figure}

\addtolength{\topmargin}{0.125in}
\addtolength{\oddsidemargin}{0.00in}


\title{Machine Learning for Precise Quantum Measurement}
\author{Alexander Hentschel}
\author{Barry C.\ Sanders\vs{5pt}}
\affiliation{Institute for Quantum Information Science, University of Calgary, Calgary, Alberta, Canada T2N 1N4}


\begin{abstract}
  \noindent
  \input{abstract_v15.2}
\end{abstract}

\maketitle 


\noindent
\input{body_v16.4}

\onecolumngrid \vs{10pt} \twocolumngrid


\vs{-13pt}


\onecolumngrid \twocolumngrid
\input SupplementalMaterial_v3

\end{document}

%% file: abstract_v15.2.tex
%
Adaptive feedback schemes are promising for quantum-enhanced measurements yet are complicated to design. Machine learning can autonomously generate algorithms in a classical setting. Here we adapt machine learning for quantum information and use our framework to generate autonomous adaptive feedback schemes for quantum measurement. In particular our approach replaces guesswork in quantum measurement by a logical, fully-automatic, programmable routine. We show that our method yields schemes that outperform the best known adaptive scheme for interferometric phase estimation.

%% file: body_v16.4.tex
%
In classical physics, it is assumed that detectors and controls can be arbitrarily accurate,
restricted only by technical limitations. 
However, this paradigm is valid only on a scale where quantum effects can be ignored.  
The `standard quantum limit' (SQL) \cite{Caves:1980rv} restricts achievable precision, 
beyond which measurement must be treated on a quantum level.
Heisenberg's uncertainty principle provides a much lower but 
insurmountable bound for the accuracy of measurement and feedback.
Approaching the Heisenberg limit is an important goal of quantum measurement.

The problem of quantum measurement can be stated as follows.
A quantity such as spatial displacement, energy fluctuation, phase shift,
or combination thereof, must be measured precisely within a specific duration of time. 
A typical device has an input and output, and the relation between the input and output
yields information from which the quantity of interest can be inferred.

Important examples of practical quantum measurement problems within limited time
include atomic clocks \cite{Bollinger.PhRvL.1985} 
and gravitational-wave detection \cite{AlexAbramovici04171992}.
Extensive efforts are underway to detect gravitational waves with laser-interferometers. 
The precision of these detectors is ultimately limited by the number of photons available 
to the interferometer within the duration of the gravitational-wave pulse \cite{Caves-Carlton-1981-PhysRevD}.
The SQL to measurement is a concern for opening up a new field
of gravitational-wave astronomy \cite{2008NatPh-Goda-etal}.

\begin{figure}[b]
	\vs{-8pt}
        \includegraphics[width=0.48\textwidth]{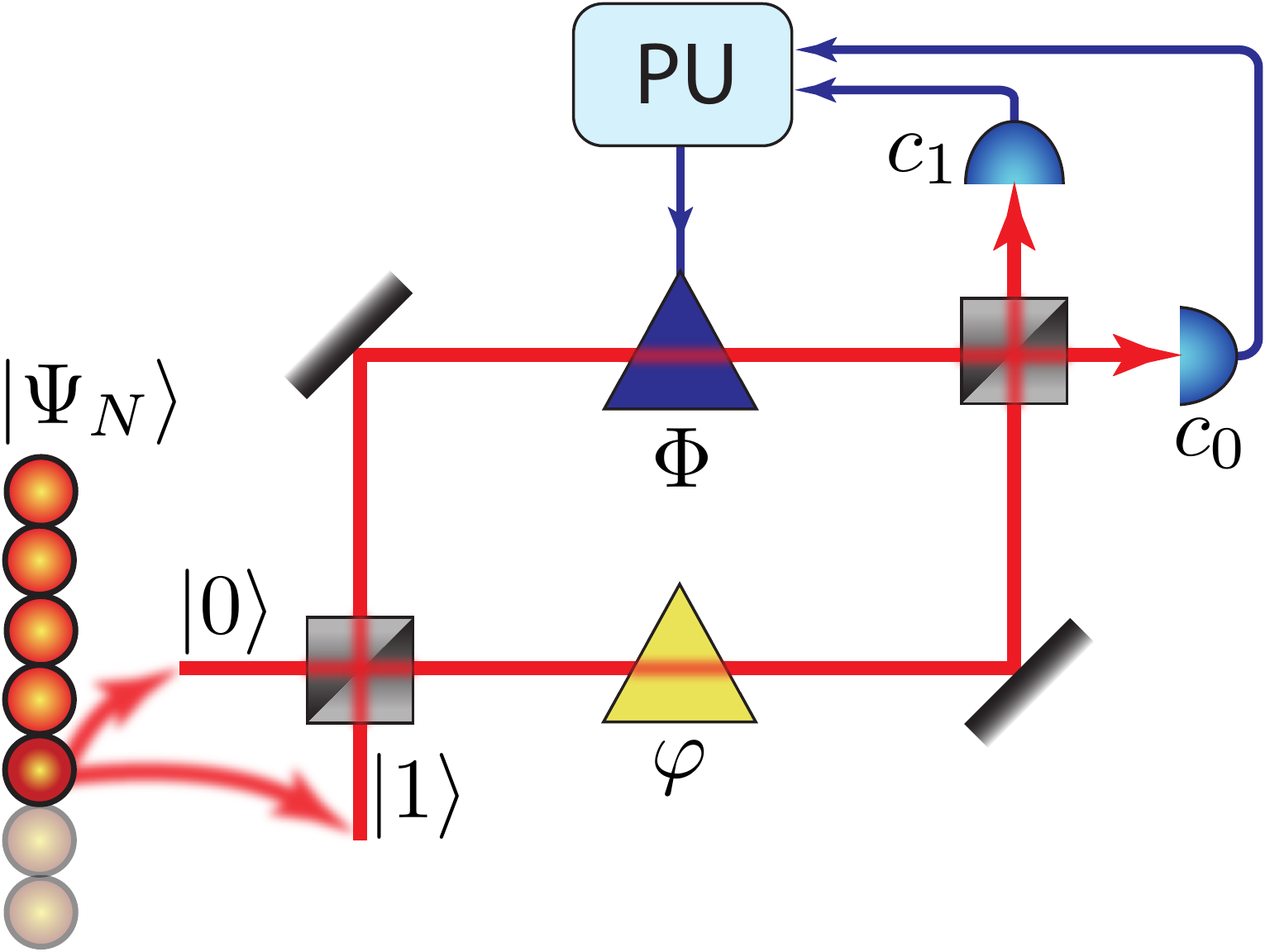}
	\vs{-15pt}
	\caption{Adaptive feedback scheme for interferometric phase estimation:~\fontseries{m}\selectfont
		Mach-Zehnder Interferometer with an unknown phase difference~$\tphase$ between the two arms and an additional controllable phase shifter $\Phi$.  
		The input state $\ket{\Psi_N}$ is stored in a quantum memory 
		and one qubit at a time is transformed into a photonic qubit and sent through the interferometer. 
		The processing unit (PU) sets the value of the phase shifter $\Phi$ depending on the measurement outcome of the single photon detectors $c_0$ and $c_1$. 
		Adaptive feedback at step $m=2$ is depicted. That is, two of the $N$ input photons (the lowest two circles) have been sent through the interferometer and measured in previous steps. 
	}
	\label{fig:MachZehnderInterferometer}
\end{figure}

For the typical two-channel interferometer, shown in Fig.~\ref{fig:MachZehnderInterferometer}, 
the goal is to estimate the relative phase shift $\tphase$ between the two channels. 
The interferometer has two input ports and two output ports, and we consider each
input and output field as being a single mode.

Each input photon to the interferometer provides a single quantum bit, or `qubit', 
as the photon is a superposition of $\ket{0}$, which represents the photon 
proceeding down one channel, or $\ket{1}$, corresponding to traversing the other channel.  
Each photon is either detected as leaving one of the two output ports or is lost.
Thus, quantum measurement can extract no more than one bit of information
about $\tphase$ per qubit in the input state \cite{Holevo_1973,cabello-2000-85}.

The fundamental precision bound is given by the `Heisenberg limit':
the standard deviation $\Delta\varphi$ of the phase estimate scales as $1/N$
for $N$ the number of input qubits used for the measurement.
$\Delta\varphi$ is determined by the error probability distribution $P(\varsigma)$ 
for estimating $\varphi$ with error $\varsigma$.
As $\varsigma$ is cyclic over $2\pi$, 
$\Delta\varphi$ is related to the Holevo variance $V$ by  \cite{Holevo1984}
\begin{equation}
	\label{eq:definition_sharpness}
	V = (\Delta\varphi)^2 =S^{-2}-1\,, \qquad S =  \left| \int_{-\pi}^{\pi} P(\varsigma) \, e^{i\varsigma} ~\text{d}\varsigma \right|.
\end{equation}
$S$ is the `sharpness' \cite{Levy-Leblond.1976} of $P(\varsigma)$.
In contrast, classical measurements only manage to 
achieve the SQL scaling $\Delta\tphase \sim 1/\sqrt{N}$
due to partition noise for photons passing through the beam splitter.
Quantum alternatives such as injecting squeezed light 
into one port of the interferometer can partially evade partition noise \cite{PhysRevLett.59.278}.

Since for any time-limited interferometric measurement, the number of input-qubits $N$ determines the achievable precision, we define $N$ as the relevant cost for the measurement.
However, it is important to discriminate resources required to operate a measurement device from the ones used to develop it.
Accordingly we distinguish between operational and developmental cost. 
The strategic question concerns the design of a device with a certain operational cost, 
so that its precision surpasses the SQL and scales as close to the Heisenberg limit as possible.

Quantum measurement schemes employing adaptive feedback are most effective, since accumulated information from measurements is exploited to maximize the information gain in subsequent measurements.
Such adaptive measurements have been experimentally shown to be a powerful technique to achieve precision beyond the standard quantum limit \cite{higgins-2007,PhysRevLett.89.133602}.
However, devising `policies', which determine feedback actions, is generally challenging and typically involves guesswork.
Our aim is to deliver a method for an automated design of policies based on machine learning 
\footnote{Per definition, a machine learning algorithm is one that has the ability to improve its performance based on past experience.}. 
To show the power of our framework, we apply it to adaptive phase estimation. As we will show, the policies generated by our method outperform the best known solutions for this problem.

Fig.\,\ref{fig:MachZehnderInterferometer} shows how a two-channel
quantum interferometric measurement with feedback operates. 
We inject a $N$-photon input state~$\left|\Psi_N\right\rangle$ into
a Mach-Zehnder interferometer with an unknown phase shift~$\varphi$
in one arm and a controllable phase shift $\Phi$ in the other arm.
Detectors at the two output ports measure which way the photon left,
and this information is transmitted to a processing unit (PU), which determines
how $\Phi$ should be adjusted for the next input-qubit.
We show that, after all $N$ input quits have been sent through the interferometer,
$\varphi$ can be inferred with a precision that scales closely to the Heisenberg limit.

We use the input $\ket{\Psi_N} = \sum_{n,k= 0}^{N} c_{n,k} \, \ket{n,N-n}$ from 
\cite{PhysRevA.63.053804}, with $c_{n,k} = (\frac{N}{2}+1)^{-\frac{1}{2}} \sin( \frac{k+ 1}{N+2}\,\pi )~\tn{e}^{\frac{i}{2}\pi(k-n)} ~ d_{n,k}^{N/2}(\frac{\pi}{2})$ and 
$ d_{\nu,\mu}^{j}(\beta)$ Wigner's (small) $d$-matrix \cite{Group_theory_and_its_application_to_QM.Wigner.1931}. $\ket{n,N-n}$ denotes a symmetrized state of $N$ suitable delayed photons with $n$ photons in channel $\ket{0}$ and $N-n$ in $\ket{1}$ \cite{Hentschel:PermutationInvariantStates}.

The challenge is to find a feedback policy, i.e.\ algorithm to run in the PU, that adjusts $\Phi$ optimally.
Fortunately, the area of machine learning suggests a promising approach.
However, standard machine learning assumes classical bits as input and output.
We inject a sequence of entangled qubits and obtain output bits. 
Due to the entanglement, the state of the remaining input qubits is progressively updated by the measurement. 
Consequently, the input  to the system (except the first qubit) depends on the unknown system parameters. %
As a result, the space of quantum measurement policies is generically non-convex, which makes policies hard to optimize.

Particle swarm optimization (PSO) algorithms \cite{Eberhart1} are remarkably successful for solving non-convex problems. 
PSO is a `collective intelligence' strategy from the field of machine learning that learns via trial and error 
and performs as well as or better than simulated annealing and genetic algorithms
\cite{SA_vs_PSO_Ethni:2009,Kennedy98matchingalgorithms,Groenwold:2002}.
Here we show that PSO algorithms also deliver automated approaches to
devising successful quantum measurement policies for implementation in the PU.

Our method is effective even if the quantum system is a black box,
i.e.\ complete ignorance about the system itself. 
The only prerequisite is a comparison criterion during the training phase
by which the success of candidate policies can be evaluated. 


To explain how we use machine learning for the quantum measurement problem,
consider the decision tree required by the PU to update the feedback $\Phi$.
The measurement of the $i^{\rm th}$ qubit yields one bit $u_i$ of information about which way the photon exited.
(If the photon is lost, there is no detection at all
 and hence no bit.  Therefore, a policy must be robust against loss.)
After $m$ photons have been processed, the PU stores the $m$-bit string 
$n_m = (u_m u_{m-1} \ldots u_1)$ and computes the feedback phase $\Phi_m$. 
In the most general case of a uniform prior distribution for $\tphase \in [0,2\pi)$, 
there is no optimal setting for the initial feedback
so we set $\Phi_0=0$, without loss of generality. All subsequent $\Phi_m$ are chosen according to a prescribed decision tree.

\begin{figure}[t]
   \includegraphics[width=0.48\textwidth]{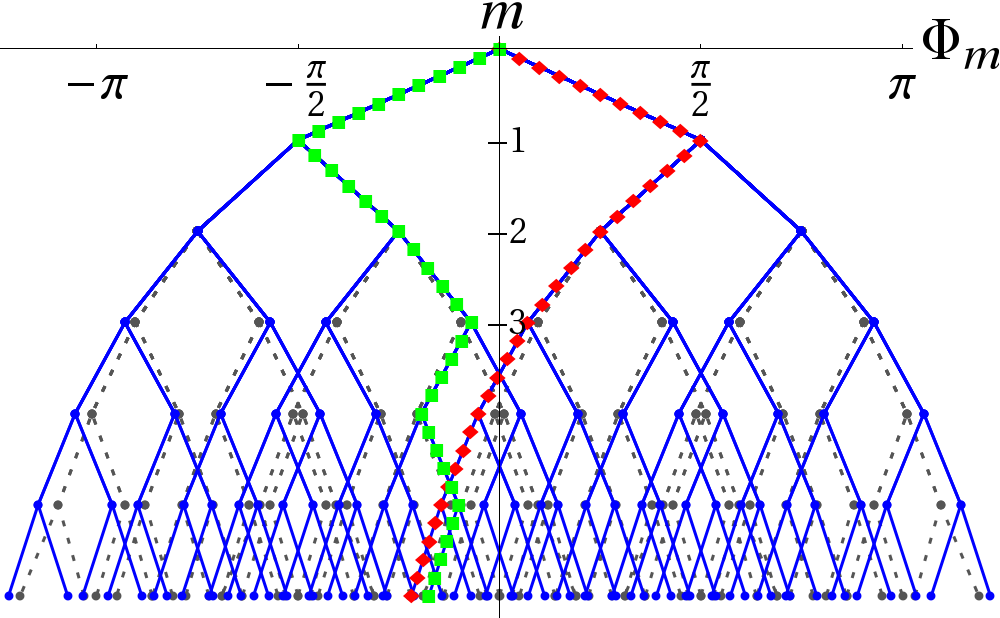}
   \vs{-17pt}
   \caption{Decision tree representations of two adaptive feedback policies for {\itshape N}$\,\mathbf{=6}$ photons.~\fontseries{m}\selectfont
		The PSO-generated policy is graphed in blue solid lines, the BWB-policy in gray dotted lines.  
		All $2^N$ possible experimental runs are represented by paths in the tree. 
		The path corresponding to an experiment with detections $u_1 u_2 \ldots u_6 = 100000$ is marked by $\textcolor{red}{\blacklozenge}$, the path corresponding to the detections 
		$u_1 u_2 \ldots u_6 = 011010$ is highlighted by $\textcolor{green}{\blacksquare}$. For each path in the tree, the inner nodes represent the applied feedback phases $\Phi_m$ 
		and the leaf shows the final phase estimate $\pEst$. 
   }
   \vs{-12pt}
   \label{fig:PSOTree}
\end{figure}

In order to show that our method not only works, but is superior to existing
feedback-based quantum measurements, we choose the Berry-Wiseman-Breslin (BWB) policy 
\cite{PhysRevA.63.053804} as a benchmark. The BWB-policy is the most precise  
policy known to date for interferometric phase estimation with direct measurement of the interferometer output.
Furthermore, its practicality has been demonstrated in a recent experiment \cite{higgins-2007}.
The BWB-policy achieves its best performance with the input state $\ket{\Psi_N}$. We use the same input state to provide fair premises. 
However, any more practical input state can be used and the PSO will autonomously learn good feedback policies.

In Fig.\,\ref{fig:PSOTree} we depict the decision trees of the BWB-policy
and of our six-photon policy.
At depth $m$, a measurement $u_{m+1}=0$ directs the path to the left and $u_{m+1}=1$ to the right.
The final destination of the path yields an estimate $\widetilde{\varphi}$
of $\varphi$, which is solely determined by the measurement record $n_N$.
Each experimental course corresponds to a path in the decision tree, 
where a path is a string of applied feedback phases 
$\Phi_0, \Phi_1(n_1), \ldots,\Phi_{N-1}(n_{N-1})$ plus a final phase estimate 
$\pEst(n_N)$.

A policy is entirely characterized by all the actions it can possibly take, 
thus by the $2^{(N+1)}-1$ phase values $\Phi_0,\, \Phi_1(0),\Phi_1(1),\, \Phi_2(00), \Phi_2(01), \ldots \in [0,2\pi)$.
Therefore, a policy can be parametrized as a vector $\rho$ in the policy space $[0,2\pi)^{2^{(N+1)}-1}$,
and any such vector $\rho$ forms a valid policy. 


For addition and scalar multiplication modulo $2\pi$, the policy space forms a vector space.
However, the dimension $2^{(N+1)}-1$ of this space grows exponentially with~$N$ making numeric optimization computationally intractable. 
Hence, we have to decrease the dimension of the search space exponentially by excluding policies. 

In the case of logarithmic search, the adjustments of the feedback phase, $\Delta\Phi_m : = | \Phi_m - \Phi_{m-1}|$, follow the recursive relation $\Delta\Phi_m = \frac{1}{2}\Delta\Phi_{m-1}$. Here, we generalize this search approach and treat $\Delta\Phi_1,\ldots,\Delta\Phi_N$ as independent variables. In the emerging trees, the adjustment $\Delta\Phi_m$ depends only on the depth $m$, i.e.\ the number of measurements performed, but not on the full measurement history $n_m$: 
\begin{equation} \label{eq:Protocol_parametrization_1}
 	\Phi_m = \Phi_{m-1} -(-1)^{u_m}\Delta\Phi_m.
\end{equation} 
Equivalently, the final phase estimate is determined via
\begin{align} \label{eq:Protocol_parametrization_2}
 	\pEst = \Phi_{N-1} -(-1)^{u_N}\Delta\tphase \, .
\end{align}
By this parametrization, $(\Delta\Phi_1,\ldots,\Delta\Phi_{N-1},\Delta\tphase)$
fully define a decision tree, because the initial feedback phase is set to $\Phi_0 =0$. 
The dimension of the resulting policy space $\mathcal{P}  = [0,2\pi)^N$ is linear in $N$.

Furthermore our dimensional reduction is promising because $\mathcal{P}$ includes a good approximation of the BWB-policy. Therefore, the best policies of this class will presumably outperform the BWB-policy. 


Now that the policy space is appropriately small, we employ a PSO algorithm. 
This population-based stochastic optimization algorithm is inspired by social 
behavior of birds flocking or fish schooling to locate feeding sites \cite{kennedy01}. 
Instead of birds and flocks, we employ the standard terms particles and swarms.

To search for the optimal phase estimation policy, the PSO algorithm models a 
swarm of particles moving in the search space $\mathcal{P}$. 
The position $\rho^{(i)} = (\Delta\Phi_1,\ldots,\Delta\Phi_{N-1},\Delta\tphase) \in \mathcal{P}$ 
of particle $i$ represents a candidate policy for estimating $\tphase$, which is initially chosen randomly.
Given the policy $\rho^{(i)}$, the sharpness $S(\rho^{(i)})$ is analytically 
computed and disclosed to the particle\SupplementMaterial{ (additional discussion in Appendix)}.

The PSO algorithm updates the candidate policies of all particles, 
i.e.\ the positions in the policy space $\mathcal{P}$, in sequential rounds.
At every time step, each particle displays the sharpest policy $g^{(i)} \in \mathcal{P}$ 
it has found so far to the rest of the swarm.
Then all particles try other policies by moving in the policy space $\mathcal{P}$. 
The moving direction for each particle is based on its own
experience and also on what other particles in its neighborhood 
have discovered is the best overall policy. 


The computation of the sharpness $S(\rho)$ has exponential time complexity in $N$. 
Consequently, policies can be optimized only for small $N$. In practice the values of $N$ achieved in experiments are quite small, much less than $14$. So small $N$ simulations are of practical value.


We have trained the quantum learning algorithm for phase estimation up to a total photon number of $N = 14$.
In each case, the PSO algorithm tries to find the sharpest policy
$(\Delta\Phi_1,\ldots,\Delta\Phi_{N-1},\Delta\tphase)$.
However, as the algorithm involves stochastic optimization,
it is not guaranteed to learn the optimal policy every time. 
So it must be run several times independently for each $N$.
Rerunning the PSO-algorithm increases the developmental cost for the policies but does not affect their operational cost.

\begin{figure}[t]
   \includegraphics[width=0.48\textwidth]{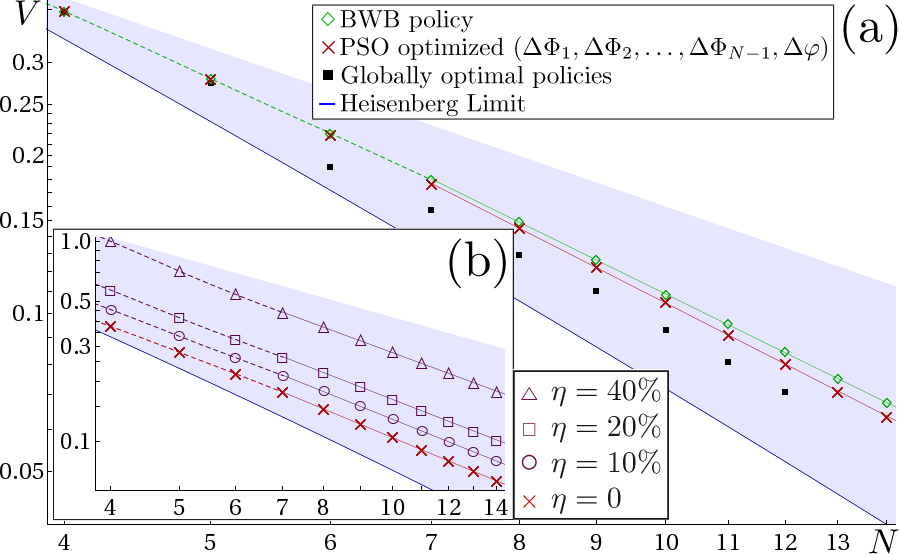}
   \vs{-17pt}
   \caption{(a) Holevo phase variance $V_\tphase$ of the PSO-optimized policies~\fontseries{m}\selectfont
        in comparison to the BWB-policy and globally optimal policies for varying operational cost $N$. 
   	The blue shaded area shows the domain of quantum enhanced measurements.~\textbf{(b)} 
	Performance of the PSO policies with probability of photon loss $\eta$. All curves follow a power law for $N \geq 7$ indicated by solid lines.
   }
   \vs{-10pt}
   \label{fig:Comparision_Variance}
\end{figure}

Fig.\,\ref{fig:Comparision_Variance}(a) depicts the performance of our quantum learning algorithm
and compares it to the BWB-policy. Within the limits of the available computational resources, 
the PSO policies outperform the BWB-policy. To provide a quantitative estimate of the performance difference, 
we calculated the scaling of the Holevo phase variance $V_\tphase$ for $N\geq 7$, 
where both curves follow a clear power law (solid lines). 
Our policy yields $V_\tphase \propto N^{-\alpha}$ with a scaling of $\alpha_\tn{PSO} = {-1.472 \pm 0.005}$, 
compared to BWB's $\alpha_\tn{BWB} = {-1.408 \pm 0.005}$.

Any practical policy has to be robust to photon loss.
In Fig.\,\ref{fig:Comparision_Variance}(b), we have graphed the performance of 
our policies for loss rates $\eta$ up to $40\%$ and calculated the scaling $\alpha_\eta$ for $N\geq 7$.  We found 
$\alpha_{0.1} = 1.421 \pm 0.006$, 
$\alpha_{0.2} = 1.377 \pm 0.008$, and
$\alpha_{0.4} = 1.307 \pm 0.009$.
This shows that our PSO-generated policies, which are optimized for a loss-less interferometer, are robust against 
moderate loss (which is also true for the BWB-policy).\SupplementMaterial{ An Extended discussion can be found in the Appendix.} Moreover, one could train the PSO algorithm for a fixed loss rate $\eta$, which increases the computation time for the sharpness evaluation by a factor of $N$.

The dimensional reduction of the search space comes at the price of possibly excluding superior policies.
In addition to proposing the BWB-policy, the authors performed in \cite{PhysRevA.63.053804} a brute force search for `globally optimal policies' in the exponential space. This was done by approximating $[0,2\pi)$ with a mesh and evaluating every possible combination of feedback phases.
The performance of the optimal policies of this search is shown in Fig.\,\ref{fig:Comparision_Variance}(a).
We found that the phase variance of the globally optimal policies is better than the performance of 
policies from our reduced space $\mathcal{P}$ only by a constant factor $0.89\pm 0.01$. 

\vs{4pt}


In summary, we have developed a framework which utilizes machine learning to autonomously 
generate adaptive feedback measurement policies for single parameter estimation problems. 
Within the limits of the available computational resources, our PSO generated policies 
achieve an optimal scaling of precision for singleshot interferometric phase estimation 
with direct measurement of the interferometer output.
Our method can be extended to allow training using a real experimental 
setup by adapting a noise tolerant PSO algorithm \cite{SWIS-CONF-2005-004}.  
This algorithm does not require prior knowledge about the physical processes involved.
Specifically, it can learn to account for all systematic experimental imperfections, 
thereby making time-consuming error modeling and extensive calibration dispensable.

\begin{acknowledgements}
\vs{4pt}
\noindent
\emph{Acknowledgments:}
We are grateful to D.\ Schuurmans, C.\ Jacob, A.\ S.\ Shastry and N.\ Khemka for intellectual contributions. We thank B. Bunk and the Humboldt-Universit\"at zu Berlin for computational resources and $i$CORE for financial support. BCS\ is a CIFAR Associate.
\end{acknowledgements}

%% file: SupplementalMaterial_v3.tex
\onecolumngrid
	\section{Appendix}
\twocolumngrid

\subsection{A. Interferometer Description}
\noindent
We use the convention 
\begin{align}
  \begin{array}{@{}ll@{}}
   	\ket{0} = \hat{a}^\dagger\vac \, , \quad
   	\ket{1} = \hat{b}^\dagger\vac \, ,
  \end{array}
\end{align}
where $\hat{a}^\dagger$ and $\hat{b}^\dagger$ are the creation operators for the field modes $a$ and $b$, and $\vac$ denotes the vacuum. 
We consider a Mach-Zehnder interferometer, where the first 50:50 beam splitter combining the two inputs has a scattering matrix 
\begin{align}
   B_1 = \frac{1}{\sqrt{2}}  \begin{pmatrix} 	1 & i \\ i & 1  \end{pmatrix}.
\end{align}
The second beam splitter $B_2$ is chosen such that it recovers the input if the phases $\Phi$ and $\tphase$ of both arms are equal, i.e. 
$B_2 =  B_1^{-1}$. The operator of the Mach-Zehnder interferometer is given by \cite{1986:Yurke-Samuel-Klauder}
\begin{align} \label{eq:MachZehnderOperator}
	\mathcal{I}(\theta) = \exp\left[ - \theta (\hat{a}^\dagger \hs{1pt} \hat{b} - \hat{a} \hs{2pt}\hat{b}^\dagger  ) \right]\, ,
\end{align}
with $\theta = \frac{1}{2}(\varphi -\Phi) $.

\subsection{B. Input States \label{sec:InputStates}}

\noindent
For single-shot interferometric phase estimation, so called `minimum uncertainty states' have been proposed to reduce the Holevo variance of the estimates of $\tphase$ \cite{PhysRevA.63.053804,PhysRevA.56.944}. 
These states are symmetric with respect to permutations of qubits and therefore the relevant quantities are the number $n_a$ and $n_b$ of photons in mode $a$ and $b$. In this case, the product of two Fock states for the two modes $a$ and $b$, denoted $\ket{n_a,n_b}$ with $N = n_a + n_b$ is a convenient basis. 

The minimum uncertainty state $\ket{\Psi_N}$ with $N$ qubits is given by
\begin{align} \label{eq:optimal_input_state}
\begin{split}
  \ket{\Psi_N}  = & \big(\minfrac{N}{2}+1 \big)^{-\frac{1}{2}} \times \\ 
  & \sum_{n,k= 0}^{N} 	\sin \hs{-2pt} \Big( \minfrac{k+ 1}{N+2}\pi \Big)
		  					e^{\frac{i}{2}\pi(k-n)} d_{n,k}^{j}(\minfrac{\pi}{2}) \ket{n,N-n}
\end{split}
\end{align}
where $ d_{\nu,\mu}^{j}(\beta)$ is Wigner's (small) $d$-matrix \cite{Group_theory_and_its_application_to_QM.Wigner.1931}.

The minimum uncertainty state has been found to be the optimal input state for single-shot adaptive interferometric phase estimation \cite{PhysRevA.63.053804}, but, due to its entanglement, it is naturally hard to prepare. 
The BWB-policy achieves its best performance under the use of the input state \eqref{eq:optimal_input_state}, we use the same input state to provide fair premises. As for the BWB-policy, any other, more practical, input state can be used and the PSO will autonomously learn a good adaptive strategy.

\subsection{C. Feedback Technique}

\noindent
The value of $\theta$ changes with the progress of the experiment due to the varying feedback $\Phi$. In our notation, $\Phi_m$ is the feedback phase applied \emph{after} the $m^\tn{th}$ detection. Hence, at the time when the $m^\tn{th}$ particle of the input state passes the interferometer, the phase difference between the two arms is parametrized by 
\begin{align}
 \theta_m := \frac{1}{2}(\varphi -\Phi_{m-1}) \, .
\end{align}
The remaining input state $\ket{\psi(n_m,\varphi)}$ after $m$ photo-detections is given by 
\begin{align} \label{eq:remaining_input_state}
  \ket{\psi(n_m,\varphi)} =  \kleinc_{u_m}(\theta_m) \cdots \kleinc_{u_2}(\theta_2)\, \kleinc_{u_1}(\theta_1)\, \ket{\Psi_N} \, ,
\end{align}
where 
\begin{align} \label{eq:latter_operator}
	\kleinc_{u_k}(\theta_k) = \frac{\hat{a}  \cos( \theta_k -u_k \minfrac{\pi}{2})- \hat{b} \sin( \theta_k -u_k \minfrac{\pi}{2})}{\sqrt{N-k+1}}
\end{align}
is the Kraus operator \cite{NielsenChuang2000} representing a measurement of the $k^\tn{th}$ particle with outcome $u_k$ \cite{PhysRevA.63.053804}.
The states \eqref{eq:remaining_input_state} are not normalized. In fact, their norm represents the probability
\begin{align} \label{eq:probability_for_measurement_n_m}
  P(n_m | \varphi) = \braket{\psi(n_m,\varphi)}{\psi(n_m,\varphi)}~.
\end{align}
for obtaining the measurement record $n_m$ given $\varphi$.

\subsection{D. Performance Measure For Policies}

\noindent
In this section, we will show how the sharpness \eqref{eq:definition_sharpness} can be analytically computed for a given policy $\rho$. Our derivation follows the procedure in \cite{PhysRevA.63.053804}.
The sharpness is determined by the probability that $\rho$ produces an estimate $\pEst_\rho$ with error $\varsigma = \pEst_\rho - \tphase$, 
\begin{align} 
  P_\rho(\varsigma | \tphase) =  \hs{-2pt} \sum_{\substack{n_N \in \\ \{0,1\}^N}} 
	\hs{-2pt} P_\rho(n_N|\tphase)~ \delta\left(\varsigma - (\pEst_\rho(n_N) - \tphase)\right)\, .
\end{align}
Here $ P_\rho(n_N|\tphase)$ is the probability that the experiment, with feedback actions determined by $\rho$, produces the measurement string $n_N$ given the phase value $\tphase$. Here we use a flat prior for $\tphase$, i.e. $P(\tphase) = \frac{1}{2\pi}$.
\begin{align}
\begin{split}
  P_\rho(\varsigma) & = \int\limits_{0}^{2\pi} P(\tphase) P_\rho(\varsigma | \tphase) \, d\tphase	\\
		      & =  \frac{1}{2\pi} \sum_{\substack{n_N \in \\ \{0,1\}^N}} P_\rho\left(n_N| \pEst_\rho(n_N) - \varsigma \right) 
\end{split}
\end{align}
From this probability distribution, we determine the sharpness with equation \eqref{eq:definition_sharpness}
\begin{align} 
  S(\rho) = ~	&\bigg|	\frac{1}{2\pi} \hs{-2pt} \sum_{\substack{n_N \in \\ \{0,1\}^N}} \int\limits_0^{2\pi} 
				 P_\rho\left(n_N| \pEst_\rho(n_N) - \varsigma \right)\,  e^{i\varsigma} \, d\varsigma  
			\bigg| 
         \notag \\ \label{eq:Agent_fitness}
		= ~	&\bigg|	\frac{1}{2\pi} \hs{-2pt} \sum_{\substack{n_N \in \\ \{0,1\}^N}} \hs{-3pt} e^{i\, \pEst_\rho(n_N)} \int\limits_0^{2\pi}  
				 P_\rho\left(n_N| \varsigma \right)\,  e^{- i\varsigma} \, d\varsigma  
			\bigg| \, .
\end{align}
The probability $P_\rho\left(n_N| \varsigma \right)$ is given by equation \eqref{eq:probability_for_measurement_n_m} and can be directly computed for a given policy $\rho$. 
(For more details see \cite{PhysRevA.63.053804}.) 
From equation \eqref{eq:Agent_fitness} it is obvious that computing the sharpness of a policy $\rho$ has complexity $\mathcal{O}(2^N)$. This is because the summand has to be evaluated for every bit-string $n_N$ of length $N$.

\subsection{E. Optimization Problem}
\noindent
Given the policy space $\mathcal{P}$, the optimization problem is defined as finding a policy 
\begin{align}
   \rho_{\rm max} \in \operatorname*{arg\,max}_{\rho \in \mathcal{P}} S(\rho)\,,
\end{align}
i.e. find a $\rho_{\rm max}$ such that $S(\rho_{\rm max}) \geq S(\rho)$ for all $\rho \in \mathcal{P}$.

\subsection{F. Details of the employed PSO algorithm}

\noindent
In this section the details of the PSO algorithm we employed are presented. The swarm $\mathcal{S} = \{p_1,p_2,\ldots,p_\swarmsize\}$ is composed of a set of particles $i = 1,2,\ldots,\swarmsize$, where  $p_i$ is the set of properties of the $i^\tn{th}$ particle and $\swarmsize \in \mathbbm{N}$ is the population size. 
At any time step $t$, $p_i$ includes the position $\rho^{(i)} \in \mathcal{P}$ of particle $i$ and $\hat{\rho}^{(i)}$, which is the best position $i$ has visited until time step $t$.  

Particle $i$ communicates with other particles in its neighborhood $\mathcal{N}^{(i)} \subseteq \mathcal{S}$. The neighborhood relations between particles are commonly represented as a graph, where each vertex corresponds to a particle in the swarm and each edge establishes a neighbor relationship between a pair of particles. This graph is commonly referred to as the swarm's population topology. 

We have adapted the common approach to set the neighborhood $\mathcal{N}^{(i)}$ of each particle in a pre-defined way regardless of the particles' position. For that purpose the particles are arranged in a ring topology. For particle $i$, all particles with a maximum distance of $r$ on the ring are in $\mathcal{N}^{(i)}$.

The PSO algorithm updates the position of all particles in a round based manner as follows. At time step $t$
\begin{enumerate}  
 \item 	Each particle $i=1,2,\ldots,\swarmsize$ assesses the sharpness $S_\varsigma(\rho^{(i)}_t)$ of its current position $\rho^{(i)}_t$ in the policy space 
	(and updates $\hat{\rho}^{(i)}$ if necessary).
 \item 	Each particle $i$ communicates the sharpest policy $\hat{\rho}^{(i)}$ it has found so far to all members of its neighborhood $\mathcal{N}^{(i)}$.
 \item 	Each particle $i$ determines the sharpest policy $ \displaystyle g^{(i)} = \max_{j \in \mathcal{N}^{(i)}} \hat{\rho}^{(j)}	$
	found so far by any one particle in its neighborhood $\mathcal{N}^{(i)}$ (including itself).
 \item 	Each particle $i$ changes its position according to 
	\begin{align} 
  	    \begin{split}
   		\rho^{(i)}_{t+1}  = ~	& \rho^{(i)}_{t} + \Delta \rho^{(i)}_t \\ 
   		\Delta \rho^{(i)}_t = ~	& \omega \,\big(\,\Delta \rho^{(i)}_{t-1}   + \varphi_1 \cdot \tn{\texttt{rand()}} \cdot(\hat{\rho}^{(i)} - \rho^{(i)}_t) 	\\ 
				& \hs{10pt} + \varphi_2 \cdot \tn{\texttt{rand()}} \cdot (g^{(i)} - \rho^{(i)}_t)  \big) 						\,.
	    \end{split}
	\end{align}
\end{enumerate}

\begin{table}[t!] 
 \begin{center}
  \begin{tabular}{ccccccccc} 
	$N$ & $\swarmsize$ & $\Delta$ 	&  $\varphi_1$ 	& $\varphi_2$  	&  $\omega$ 	& $\nu_{\tn{max}}$ & $r$ 	&	$\lambda$	\\
	  4 	&  50	   & 700  	&   0.5	 	&   1		&  1		& 0.05		   &  1		&	100\%		\\
	  5 	&  50	   & 700  	&   0.5	 	&   1		&  1		& 0.05		   &  1		&	100\%		\\
	  6 	&  50	   & 700  	&   0.5	 	&   1		&  1		& 0.05		   &  1		&	100\%		\\
	  7 	&  50	   & 500  	&   0.5 	&   1 		&  0.8		& 0.2		   &  4		&	100\%		\\
	  8 	&  60	   & 300  	&   0.5 	&   1 		&  0.8		& 0.2		   &  6		&	35\%		\\
	  9 	&  60	   & 500  	&   0.5 	&   1 		&  0.8		& 0.2		   &  6		&	33\%		\\
	  10 	&  60	   & 400  	&   0.5 	&   1 		&  0.8		& 0.2		   &  6		&	25\%		\\
	  11 	&  60	   & 400  	&   0.5 	&   1 		&  0.8		& 0.2		   &  6		&	66\%		\\
	  12 	&  120	   & 1000 	&   0.5 	&   1 		&  0.8		& 0.2		   &  12	&	20\%		\\
	  13 	&  375	   & 300  	&   0.5 	&   1 		&  0.8		& 0.2		   &  30	&	17\%		\\
	  14 	&  441	   & 100  	&   0.5 	&   1 		&  0.8		& 0.2		   &  35	&	20\%		\\
  \end{tabular} 
  \caption{PSO settings for $N$-Photon input state: velocity damping $\omega$, swarm size $\swarmsize$, number of PSO-Steps $\Delta$, 
	   max stepsize $\nu_{\tn{max}}$, exploitation weight $\varphi_1$, exploration weight $\varphi_2$, fraction $\lambda$ of PSO runs produced policies with variance depicted in Fig.\ \ref{fig:Comparision_Variance}.} 
  \label{tab:PSO_settings}
\end{center}
\vs{-10pt}
\end{table}

\begin{table*}[t!]
 \begin{center}
  \begin{tabular}{clcc} 
	$N$ 	& $\Delta\Phi_1,\ldots,\Delta\Phi_{N-1}$ &  $\Delta\tphase$ & $V_\tphase$  \\
 	4 	& $1.5701, 0.7862, 0.5043$ 													&  $0.3507$ & 0.37621 \\
 	5 	& $1.5722, 0.7816, 0.5293, 0.3684$ 												&  $0.2739$ & 0.27922 \\
 	6 	& $1.5708, 0.7830, 0.5669, 0.3881, 0.2889$ 											&  $0.2306$ & 0.21835 \\
 	7 	& $1.5708, 0.7854, 0.6159, 0.4130, 0.3073, 0.2421$ 										&  $0.1988$ & 0.17630 \\
 	8 	& $1.5708, 0.7854, 0.6663, 0.4399, 0.3264, 0.2551, 0.2080$									&  $0.1750$ & 0.14561 \\
 	9 	& $1.5708, 0.7854, 0.7079, 0.4620, 0.3440, 0.2671, 0.2164, 0.1811$ 								&  $0.1554$ & 0.12253 \\
 	10 	& $1.5708, 0.7854, 0.7392, 0.4788, 0.3599, 0.2780, 0.2240, 0.1867, 0.1597$ 							&  $0.1393$ & 0.10482 \\
 	11 	& $1.5706, 0.7850, 0.7613, 0.4934, 0.3744, 0.2875, 0.2313, 0.1920, 0.1642, 0.1421$ 						&  $0.1260$ & 0.09094 \\
 	12 	& $1.5708, 0.7854, 0.7800, 0.5023, 0.3890, 0.2983, 0.2384, 0.1973, 0.1677, 0.1456, 0.1285$ 					&  $0.1149$ & 0.07985 \\
 	13 	& $1.5695, 0.7847, 0.7920, 0.5119, 0.4029, 0.3083, 0.2457, 0.2027, 0.1720, 0.1487, 0.1310, 0.1170$ 				&  $0.1054$ & 0.07083 \\
	14 	& $1.5703, 0.7860, 0.8018, 0.5195, 0.4171, 0.3179, 0.2529, 0.2077, 0.1756, 0.1517, 0.1335, 0.1190, 0.107326$			&  $0.0975$ & 0.06337 \\
  \end{tabular} 
  \caption{Parameters for the best PSO-generated policies for $N=4,\ldots,14$.}
  \label{tab:PSO_optimized_Feedback_strategies}
\end{center}
\vs{-10pt}
\end{table*}

The parameter $\omega$ represents a damping factor that assists convergence, and $\verb|rand()|$ is a function returning uniformly distributed random numbers in $[0,1]$.
The `exploitation weight' $\varphi_1$ parametrizes the attraction of a particle to its best previous position $\hat{\rho}^{(i)}$,
and the `exploration weight' $\varphi_2$ describes the attraction to the best position $g^{(i)}$ in the neighborhood. To increase convergence, we bound each component of $\Delta \rho^{(i)}_s$ by a maximum value of $\nu_{\tn{max}}$. In sumary, the properties of the swarm, such as size and behavior, are defined by the following parameters.
\begin{align}\label{list:pso_inpit_parameters}
{
	\begin{array}{@{\hs{-2pt}}rl@{}}
	   \omega \in [0,1] 								& \tn{velocity damping factor} 					\vs{1.5pt}\\
	   \varphi_1 \in [0,1] 								& \tn{exploitation weight} 					\vs{1.5pt}\\
	   \varphi_2  \in [0,1]								& \tn{exploration weight} 					\vs{1.5pt}\\
	   \swarmsize									& \tn{population size}						\vs{1.5pt}\\
	   \nu_{\tn{max}}								& \begin{array}{@{}l@{}}
												\tn{maximum step size } 		\vs{-2.5pt}\\ 
												\tn{particles are allowed to move}
				  							  \end{array} 							\vs{1.5pt}\\
	   r										& \tn{interaction range of particles}
	\end{array}
}
\end{align}
Clearly, the success and the number of required PSO steps to find the maximum is highly dependent on the values of these parameters. For instance, with increasing $N$, a bigger population size is required to account for the raising dimensionality of the search space. The most successful settings we found are listed in Table \ref{tab:PSO_settings}.
For each $N=1,\ldots,14$, the best policy $(\Delta\Phi_1,\ldots,\Delta\Phi_{N-1},\Delta\tphase)$ the PSO algorithm learned is given Table \ref{tab:PSO_optimized_Feedback_strategies}.

\subsection{G. Noise resistance}
\noindent
As with the BWB-policy, which works with an idealized noiseless model, the training of our PSO algorithm is performed based on the simulation of a noiseless Mach-Zehnder interferometer. However, Higgins et al. recently used the BWB-policy as a component for their experiment \cite{higgins-2007}, which shows that the feedback policies we considered for optimization are robust against noise. 

Therefore, the policies generated by our learning approach are applicable to moderately noisy experiments, even though the PSO algorithm was trained on a simulated noisefree experiment.

Figure \ref{fig:Comparision_Variance} shows the Holevo variance of the PSO-generated policies for different photon-loss rates. We have calculated the variance as follows. For a fixed loss-rate $\eta$, the probability of detecting $k$ of the $N$ input-photons is given by the binomial distribution 
$B(k;N,\eta) = \big(\hs{-1pt}\atop{N}{k}\hs{-1pt}\big) \eta^{(N-k)}(1-\eta)^k$. Then the probability that the policy $\rho$ produces an estimate $\pEst_\rho$ with error $\varsigma = \pEst_\rho - \tphase$ is 
\begin{align}
   P_\rho(\varsigma | \varphi) = \sum_{k=0}^{N} B(k;N,\eta) P(\varsigma | \varphi, k)\,.
\end{align}
An analogous calculation to the one in appendix D yields
\begin{align} \label{eq:Sharpness_with_loss}
  S(\rho) &= ~	\bigg|	\sum_{k=0}^{N} B(k;N,\eta) \mathscr{S}(k) \bigg| 
\end{align}
with 
\begin{align}
\mathscr{S}(k) &= \frac{1}{2\pi} \hs{-2pt} \sum_{\substack{n_k \in \\ \{0,1\}^k}} \hs{-3pt} e^{i\, \pEst_\rho(n_k)} \int\limits_0^{2\pi}  
				 P_\rho\left(n_k| \varsigma \right)\,  e^{- i\varsigma} \, d\varsigma  \,.
\end{align}
The probability $P_\rho\left(n_k| \varsigma \right)$ is given by equation \eqref{eq:probability_for_measurement_n_m}.

Figure \ref{fig:Comparision_Variance} shows that our PSO-generated policies with the state \eqref{eq:optimal_input_state} as input are remarkably robust against photon loss. Even with a loss rate of $\eta = 40\%$ the variance scales as $N \propto N^{-1.307 \pm 0.009}$, and the measurement lies in the domain of quantum enhanced measurements. 

The strong robustness against photon loss is mainly due to the nature of the input state \eqref{eq:probability_for_measurement_n_m}, which is highly entangled and symmetric with respect to qubit permutations. As a consequence, this state remains entangled even if a high percentage of photons are lost.